\documentclass[aps,pre,onecolumn,notitlepage,footinbib,longbibliography]{revtex4-1}
\pdfoutput=1
\usepackage{amsmath}
\usepackage{amsfonts}
\usepackage{amssymb}
\usepackage{lmodern,dsfont}
\usepackage{graphicx}
\usepackage[usenames,dvipsnames]{xcolor}
\usepackage{bm}
\usepackage{blkarray}
\usepackage{blindtext}
\usepackage[english=american]{csquotes}

\newcommand{\Av}[1]{\left\langle #1 \right\rangle}
\newcommand{\av}[1]{\langle #1 \rangle}
\newcommand{\n}{\nonumber}
\newcommand{\nn}{\nonumber \\}

\begin{document}

\author{Andreas Dechant}
\affiliation{Department of Physics \#1, Graduate School of Science, Kyoto University, Kyoto 606-8502, Japan}
\affiliation{WPI-Advanced Institute of Materials Research (WPI-AIMR), Tohoku University, Sendai 980-8577, Japan}
\author{Shin-ichi Sasa}
\affiliation{Department of Physics \#1, Graduate School of Science, Kyoto University, Kyoto 606-8502, Japan}
\title{Current fluctuations and transport efficiency for general Langevin systems}
\date{\today}

\begin{abstract}
We derive a universal bound on generalized currents in Langevin systems in terms of the mean-square fluctuations of the current and the total entropy production.
This bound generalizes a relation previously found by Barato et al. to arbitrary times and transient states.
Using the bound, we define a new efficiency for stochastic transport, which measures how close a given system comes to saturating the bound.
The existence of such a bounded efficiency implies that stochastic transport is unavoidably accompanied by a fluctuations and dissipation, which cannot be reduced arbitrarily.
We apply the definition of transport efficiency to steady state particle transport and heat engines and show that the transport efficiency may approach unity at finite current, in contrast to the thermodynamic efficiency.
Finally, we derive a bound on purely diffusive transport in terms of the Shannon entropy.
\end{abstract}

\maketitle

\section{Introduction}
In microscopic systems, transport is almost universally stochastic. 
Due to the small length and energy scales, noise inevitably leads to diffusion and any observed current will have non-negligible fluctuations.
Such stochastic currents play a crucial role for the motion of cells or transport of chemicals inside a cell \cite{Lin77,Neu92,Mer01} and many more systems in biology, physics and beyond \cite{Sch73,Bee91,Pra02,Der04,Ber05,San07,Ant08,Cil10}.
In all these situations one has a competition between transport, diffusion and the energy input necessary to sustain the transport, the goal generally being to maximize the former while minimizing the latter two \cite{Wan02,Suz03,Mac04}.
If the transport is driven thermally, i.~e.~by a temperature difference, then the thermodynamic efficiency -- the ratio of the work performed during the transport and the energy absorbed from the heat bath -- is bounded by the Carnot efficiency \cite{Fey65}.
Any work extraction is accompanied by dissipation, which increases the entropy of the heat bath, leading to this universal upper bound.

The thermodynamic efficiency may, however, not be very useful as a measure of the efficiency of the transport.
First, in many cases, the utility is the transport itself rather than the performed work.
Second, even though the transport may be efficient in a thermodynamic sense, it may be accompanied by a large diffusion, making the transport unreliable and thus not useful in practice \cite{Mac04,Kri05}.
For these reasons, transport is often characterized in terms of the Pecl{\'e}t number \cite{Peu09}, which is defined as the current times the system size divided by the diffusivity.
However, this does not take into account the cost of generating the current and thus the Pecl{\'e}t number may in principle be arbitrarily large.
By contrast, the so-called Stokes-efficiency $\eta^\text{s}$ introduced in Ref.~\cite{Bus01,Wan02}, related to the rectification efficiency \cite{Suz03,Mac04}, is defined as the square of the current divided by the rate of free energy consumption. 
It takes into account the cost of maintaining the current and is bounded $\eta^\text{s} \leq 1$, but neglects diffusion.
Thus, a bounded measure of transport efficiency relating the current to both fluctuations and dissipation is desirable.

In this Letter, we study the fluctuations of a current $\dot{R}(t)$ in general out-of-equilibrium Langevin dynamics.
We derive a very general bound on the cumulant generating function of this current, generalizing the previous result obtained by Nemoto et al.~\cite{Nem11} for the long-time limit of steady state dynamics to time-dependent dynamics and arbitrary times.
In the absence of time-dependent driving, we show that this bound yields the equivalent of the thermodynamic uncertainty relation introduced by Barato et al. \cite{Bar15} and later generalized to finite-time fluctuations of the current \cite{Pie17,Hor17}.
However, in contrast to previous results, our proof does not rely on any large deviation arguments and is not limited to the steady state.
This allows us to define a transport efficiency $\chi^R$---the ratio of the average current squared, divided by the product of the time-dependent diffusivity and the time-averaged entropy production---which is universally bounded $\chi^R \leq 1$.
The existence of such an efficiency necessitates a tradeoff between the achieved current on the one hand, and the fluctuations of the current and the cost of driving on the other hand.
For a particle diffusing in a periodic potential and periodic temperature profile, a ratchet-type model first studied by B{\"u}ttiker and Landauer \cite{Bue87,Lan88}, we show that the bound may be saturated at finite current.
We further use the bound on the transport efficiency for work currents to derive a trade-off relation between power and thermodynamic efficiency of steady state heat engines, complementary to the time-periodic engines studied in Ref.~\cite{Shi16}.
Finally, we discuss a diffusive relaxation process in which the current is driven by a difference in Shannon entropy.
The bound on the transport efficiency then allows us to estimate the current in terms of the geometry of the system.

\section{Bound on the generating function} 
We consider a system consisting of $M$ interacting degrees of freedom represented by the state vector $\bm{x} = \lbrace x_1, x_2, \ldots, x_M \rbrace$, whose time-evolution $\bm{x}(t)$ is described by the set of coupled Langevin equations 
\begin{align}
\dot{\bm{x}}(t) = \bm{A}(\bm{x}(t),t) + \sqrt{2} \bm{G}(\bm{x}(t),t) \cdot \bm{\xi}(t) \label{langevin},
\end{align} 
where $\bm{A}(\bm{x},t)$ is the drift vector, corresponding to systematic generalized forces, and $\bm{G}(\bm{x},t)$ is a full-rank $M \times M$ matrix.
The source of randomness is the Gaussian white noise vector $\bm{\xi}$, whose components we take to be mutually independent $\av{\xi_i(t) \xi_j(s)} = \delta_{i j} \delta(t-s)$.
Since the matrix $\bm{G}(\bm{x},t)$ may depend on the state of the system, the noise is generally multiplicative and we adopt the Ito-convention.
We assume that the initial state of the system is sampled from a prescribed distribution $P_0(\bm{x})$.
We can equivalently describe the system in terms of its probability density and the Fokker-Planck equation 
\begin{subequations}
\begin{align}
\partial_t P(\bm{x},t) &= -\bm{\nabla} \bm{J}(\bm{x},t) \label{continuity} \\
\bm{J}(\bm{x},t) &= \big(\bm{A}(\bm{x},t) - \bm{\nabla} \bm{B}(\bm{x},t) \big) P(\bm{x},t) \label{current} ,
\end{align}\label{fokkerplanck}%
\end{subequations}
with initial condition $P(\bm{x},0) = P_0(\bm{x})$.
Here we introduced the symmetric diffusion matrix $\bm{B}(\bm{x},t) = \bm{G}(\bm{x},t) \bm{G}^T(\bm{x},t)$, where $T$ denotes transposition.
Our goal is to study currents induced by the dynamics Eq.~\eqref{langevin}.
To this end, we define a generalized time-integrated current $R(t)$ via its time derivative \cite{Che15}
\begin{align}
\dot{R}(t) = \bm{Z}(\bm{x}(t),t) \circ \dot{\bm{x}}(t) \label{generalized-current},
\end{align}
with the Stratonovich product \enquote{$\circ$}.
Such a current could be the displacement of a particle, but also the heat exchanged with a heat bath or the fluctuating entropy production.
In order to study the fluctuations of $R(t)$, we consider the scaled cumulant generating function \cite{Nem11}
\begin{align}
K_R(h,\mathcal{T}) &= \frac{1}{\mathcal{T}} \ln \Av{e^{h \int_0^\mathcal{T} \text{d}t \ \dot{R}(t)}} \label{cgf} ,
\end{align}
where $\av{\ldots}$ denotes an average over the noise history.
The central mathematical result of this Letter is the following bound on the scaled cumulant generating function:
\begin{align}
K_R(h,\mathcal{T}) \geq \frac{1}{\mathcal{T}}\int_0^\mathcal{T} \text{d}t \ \bigg( h \av{\dot{R}}_t^Y - \frac{1}{4} \av{\bm{Y} \bm{B}^{-1} \bm{Y}}_t^Y \bigg) \label{variational},
\end{align} 
where the averages $\av{\ldots}_t^Y$ are taken with respect to the solution of the modified Fokker-Plack equation
\begin{align}
\partial_t P^Y(\bm{x},&t) = -\bm{\nabla} \bm{J}^Y(\bm{x},t) \label{fokkerplanck-mod} \\
\bm{J}^Y(\bm{x},t) &= \Big(\bm{A}(\bm{x},t) + \bm{Y}(\bm{x},t)  - \bm{\nabla} \bm{B}(\bm{x},t) \Big) P^Y(\bm{x},t) \n.
\end{align}
Thus any set of added generalized forces $\bm{Y}(\bm{x},t)$ yields a lower bound on the generating function.
We give a straightforward proof of the lower bound towards the end of this Letter.
At this point, it is natural to ask whether an optimal set of forces maximizing the bound exists, and if so, whether the inequality in Eq.~\eqref{variational} may become an equality, yielding a variational formula for the generating function as derived in Ref.~\cite{Nem11}.
We show in Appendix \ref{app-optimal} that such a set of optimal forces exists, corresponding to an optimal tilting of the stochastic process \cite{Har12,Che15b}. 
However, for the following results, the bound itself is sufficient.

\section{Thermodynamic uncertainty relation and transport efficiency} 
Since every choice for the generalized forces $\bm{Y}(\bm{x},t)$ yields a lower bound on the generating function, the task is now to find a good choice, which leads to a physically meaningful bound.
In the following, we consider the case where the drift vector $\bm{A}$ and diffusion matrix $\bm{B}$ have no explicit time-dependence.
In this case, a time-rescaling of the original process, $P^Y(\bm{x},t) = P(\bm{x},(1+\alpha)t)$, $\bm{J}^Y(\bm{x},t) = (1+\alpha) \bm{J}(\bm{x},(1+\alpha)t)$, can be realized by choosing
\begin{align}
\bm{Y}(\bm{x},t) = \alpha \frac{\bm{J}(\bm{x},(1+\alpha)t)}{P(\bm{x},(1+\alpha)t)} ,
\end{align}
for any $\alpha > -1$.
Provided that the variables $\bm{x}$ are even under time-reversal, and thus the current $\bm{J}$ is an irreversible current, the resulting bound can be written as
\begin{align}
K_R(h,\mathcal{T}) \geq \frac{1}{\mathcal{T}} \bigg( h \av{R}_{(1+\alpha)\mathcal{T}} - \frac{\alpha^2 \av{\Sigma}_{(1+\alpha)\mathcal{T}}}{4 (1+\alpha)} \bigg)
\end{align}
where $\av{\Sigma}_{\mathcal{T}} = \int_0^\mathcal{T} \text{d}t \ \av{\bm{J} \bm{B}^{-1} \bm{J} / P^2}_t$ is the total entropy production \cite{Spi12}.
The generating function at time $\mathcal{T}$ is thus related to the current and entropy production at the rescaled time $(1+\alpha)\mathcal{T}$.
We now choose the rescaling parameter as $\alpha = 2 h \mathcal{T} \langle\dot{R}\rangle_\mathcal{T}/\langle\Sigma\rangle_{\mathcal{T}}$ and expand the right-hand side for small $h$,
\begin{align}
K_R(h,\mathcal{T}) \geq \frac{h}{\mathcal{T}} \bigg( \av{R}_\mathcal{T} + h \frac{\big( \mathcal{T} \av{\dot{R}}_\mathcal{T} \big)^2}{\av{\Sigma}_{\mathcal{T}}} \bigg) + O(h^3) \label{cgf-bound}.
\end{align}
If the system is initially in the steady state, $\langle R \rangle_\mathcal{T} = \langle\dot{R}\rangle^\text{st} \mathcal{T}$ and $\langle\Sigma\rangle_\mathcal{T} = \sigma_\text{m} \mathcal{T}$, with the rate of entropy production in the medium $\sigma_\text{m}$, then the higher order terms vanish and we recover the bound on the finite time generating function that was recently proposed for continuous-time Markov jump processes in Ref.~\cite{Pie17} and subsequently proven in Ref.~\cite{Hor17}.
Our analysis provides the proof of this bound for Langevin dynamics, without requiring any large deviation arguments.
However, Eq.~\eqref{cgf-bound} is more general since it is also applicable to relaxational dynamics, where the system is not in the steady state.
Expanding the generating function up to second order in $h$, we get a bound on the instantaneous current
\begin{align}
\av{\dot{R}}_\mathcal{T}^2 \leq D_{R,\mathcal{T}} \ \bar{\sigma}_\mathcal{T} \label{uncertainty},
\end{align}
where we defined the time-dependent diffusivity $D_{R,\mathcal{T}} = \langle \Delta R^2\rangle_\mathcal{T}/(2 \mathcal{T})$, with the mean-square fluctuations $\langle\Delta R^2\rangle_\mathcal{T} = \langle R^2\rangle_\mathcal{T} - \langle R\rangle_{\mathcal{T}}^2$, and the time-averaged entropy production rate $\bar{\sigma}_\mathcal{T} = \langle \Sigma\rangle_\mathcal{T}/\mathcal{T}$.
This relation is the second main result of this Letter.
This bound constitutes the extension of the thermodynamic uncertainty relation introduced in Ref.~\cite{Bar15} to finite times and systems out of steady state.
If the generalized current is the fluctuating entropy production, this result recovers the Fano-factor inequality proven in \cite{Pig17}.
In the following, we will explore the physical consequences of this inequality for stochastic transport in general and in three simple but representative examples in particular.
Before we do so, let us remark on a particularity of the bound \eqref{uncertainty} as presented here.
Note that the quantity $\av{\dot{R}}_\mathcal{T}$ is depends only on the probability current at time $\mathcal{T}$
\begin{align}
\av{\dot{R}}_\mathcal{T} = \int \text{d}\bm{x} \ \bm{Z}(\bm{x},\mathcal{T}) \bm{J}(\bm{x},\mathcal{T})
\end{align}
and is the instantaneous current at time $\mathcal{T}$.
Thus, the left-hand side of Eq.~\eqref{uncertainty} is an instantaneous quantity at time $\mathcal{T}$, while the quantities on the right-hand side depend on the entire process from time $0$ up to time $\mathcal{T}$.
That such a relation between instantaneous and time-integrated currents exists is a consequence of the explicit time-independence of the dynamics (i.~e.~the drift vector and diffusion matrix).
It implies that the currents in an overdamped relaxational dynamics are always well-behaved in the sense that any sudden changes in the current have to be preceded by a large entropy production.

A direct consequence of Eq.~\eqref{uncertainty} is the existence of a transport efficiency
\begin{align}
\chi_R \equiv \frac{\av{\dot{R}}_\mathcal{T}^2}{D_{R,\mathcal{T}} \ \bar{\sigma}_\mathcal{T}} \leq 1 \label{transport-efficiency} .
\end{align}
This efficiency relates the desired effect of stochastic transport (the current) to the cost (dissipation in the form of increasing entropy) and undesired side effects (fluctuations of the current).
For any Langevin dynamics without explicitly time-dependent drift vector and diffusion matrix, this efficiency is bounded by $1$ at arbitrary times and independent of the initial state.
Thus, one cannot obtain an arbitrary large current without either accepting large fluctuations or investing a large cost in terms of entropy production to drive the current.
Note that the entropy production rate appearing in Eq.~\eqref{transport-efficiency} is the total entropy production, which includes both the heat dissipated into the environment and the difference in Shannon entropy \cite{Gib02,Sha48} between initial and final state.

\section{Applications}

\subsection{Steady-state transport}
\begin{figure}
\includegraphics[width=.47\textwidth]{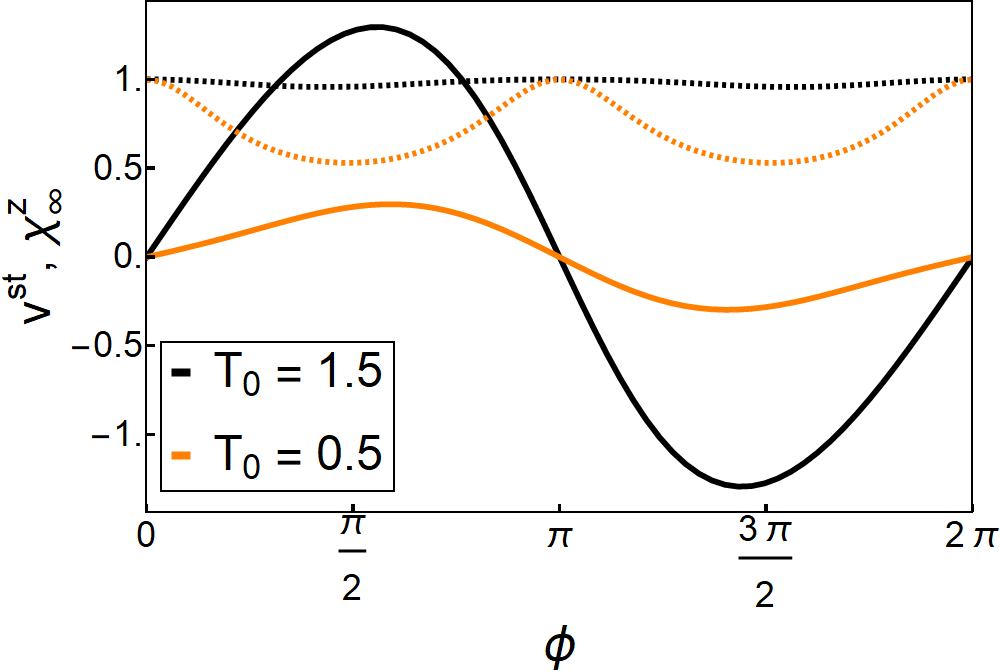}
\caption{(Color online) Steady state current and transport efficiency for the B{\"u}ttiker-Landauer ratchet at different temperatures as a function of the phase shift between potential and temperature profile. In both cases, the transport efficiency $\chi^z_\infty$ (Eq.~\eqref{LB-transport-efficiency}, dashed lines) is strictly $1$ only for vanishing  current $v^\text{st} = 0$ (solid lines). However, for higher temperatures (black) the transport efficiency approaches $1$ even at finite current, indicating that the bound can be saturated. The remaining parameter values are $U_0 = 1$, $L = 1$ and $\theta = 1/2$. \label{fig-1}}
\end{figure}

Consider an overdamped Brownian particle moving in a one-dimensional periodic potential $U(x+L) = U(x)$ and in a spatially varying temperature field $T(x+L) = T(x)$, where $x$ denotes the position of the particle.
The motion of the particle can be described by the Langevin equation
\begin{align}
\gamma \dot{x}(t) = -U'(x(t)) + \sqrt{2 \gamma T(x(t))} \cdot \xi(t) \label{LB-langevin},
\end{align}
where $\gamma$ is the damping coefficient of the medium, $\xi(t)$ is Gaussian white noise $\av{\xi(t) \xi(t')} = \delta(t-t')$ and $\cdot$ denotes the It{\=o}-product.
We set the mass of the particle and the Boltzmann constant to $1$ without loss of generality.
As was shown by B{\"u}ttiker and Landauer \cite{Bue87,Lan88}, this system generally exhibits a non-zero steady state velocity $v^\text{st}$, the displacement $z(t) = x(t) - x(0)$ growing on average as $\av{z(t)} = v^\text{st} t$.
The system thus operates as a ratchet, converting the unbiased fluctuations of the noise into a directed current.
We identify the generalized current corresponding to the motion of the particle as $\dot{z}(t) = \dot{x}(t)$.
At the same time, the displacement of the particle fluctuates, quantified by the mean-square displacement $\langle \Delta z^2 \rangle_\mathcal{T}$.
The transport efficiency Eq.~\eqref{transport-efficiency} for this system in the steady state is then given by
\begin{align}
\chi_{z,\mathcal{T}} =  \frac{{v^\text{st}}^2}{D_{z,\mathcal{T}} \ \sigma_\text{m}} \leq 1 . \label{LB-transport-efficiency}
\end{align}
Note that the effective diffusivity $D_{z,\mathcal{T}}$ depends on time but converges towards a unique long-time value $D_{z,\infty}$.
Since all other quantities are independent of time, we may bound the long-time asymptotic transport efficiency by $\chi_{z,\infty} \leq (v_\text{st})^2/(D_{z,\text{min}} \sigma_\text{m}) \leq 1$, where $D_{z,\text{min}}$ is the minimal value of $D_{z,\mathcal{T}}$.
Thus reaching a transport efficiency of unity at long times requires $D_{z,\infty} = D_{z,\text{min}}$.
We provide explicit expressions for the steady state velocity $v^\text{st}$, long-time diffusivity $D_{z,\infty}$ and entropy production rate $\sigma_\text{m}$ in Appendix \ref{app-blr}.
As a concrete example, we consider the sine-shaped potential $U(x) = U_0 \sin(2 \pi x/L)$ and temperature profile $T(x) = T_0 (1 + \theta \sin(2 \pi x/L + \phi))$ with $\theta \leq 1$.
The resulting current and transport efficiency are shown in Fig.~\ref{fig-1}.
While we have $\chi_{z,\infty} = 1$ only for vanishing current (and zero dissipation), the transport efficiency for this model generally increases with the overall temperature $T_0$.
In particular for $T_0 \gg U_0$, the transport efficiency can be arbitrarily close to $1$ while the current remains finite.

\subsection{Steady state heat engines}
By applying a constant load force $F_0$ to the ratchet discussed above, it can perform work against the external force at a rate $\dot{W}(t) = - F_0 \dot{x}(t)$.
More generally, we consider a system connected to two heat baths at temperatures $T_\text{h} > T_\text{c}$.
For the B{\"u}ttiker-Landauer ratchet this corresponds to parameterizing the temperature profile as $1/T(x) = 1/T_\text{c} + (1/T_\text{h} - 1/T_\text{c}) \psi(x)$ with some function $0 \leq \psi(x) \leq 1$.
In this case, we define the heat exchange with the hot and cold bath analog to the definition for a time-dependent temperature in Ref.~\cite{Bra14},
\begin{align}
\av{\dot{Q}_\text{h}} &= -\frac{1}{\gamma} \int_0^L \text{d}x \ \psi(x) \Big( V'(x)^2 - T(x) V''(x) \Big) P^\text{st}(x) \\  \av{\dot{Q}_\text{c}} &= -\frac{1}{\gamma} \int_0^L \text{d}x \ \big(1 - \psi(x)\big) \Big( V'(x)^2 - T(x) V''(x) \Big) P^\text{st}(x) \n,
\end{align}
with $V(x) = U(x) + T(x)$, see Eq.~\eqref{sigma-bl}.
This definition ensures that only regions with $T(x) > T_\text{c}$ contribute to $Q_\text{h}$ and vice versa.
Another widely known example for a steady state heat engine is the Feynman-Smoluchowski ratchet \cite{Smo12,Fey65,Sek97}.
In both cases, in the steady state, the entropy production rate is
\begin{align}
\sigma^\text{m} &= -\frac{\langle\dot{Q}_\text{c}\rangle}{T_\text{c}}  - \frac{\langle\dot{Q}_\text{h}\rangle}{T_\text{h}}  = \frac{\langle\dot{W}\rangle}{T_\text{c}} \bigg( \frac{\eta_\text{C}}{\eta} - 1 \bigg) ,
\end{align}
where $\dot{Q}_\text{h/c}$ is the rate at which the system exchanges heat with the hot/cold heat bath and we used the first law  $\langle \dot{W} \rangle = \langle\dot{Q}_\text{c}\rangle + \langle\dot{Q}_\text{h}\rangle$,
defined the thermodynamic efficiency for $\langle \dot{W} \rangle \geq 0$ as $\eta = \langle \dot{W} \rangle/\langle\dot{Q}_\text{h}\rangle$ and introduced the Carnot efficiency $\eta_\text{C} = 1- T_\text{c}/T_\text{h}$. We then express the transport efficiency for work as
\begin{align}
\chi_\infty^W = \frac{\langle \dot{W} \rangle^2}{D_{W,\infty} \ \sigma_\text{m}} = \eta T_\text{c} \frac{\av{\dot{W}}}{D_{W,\infty} (\eta_\text{C} - \eta)} .
\end{align}
The bound $\chi_{W,\infty} \leq 1$ on the transport efficiency implies that that extracting work from the system not only requires a finite dissipation---as per the Second Law---but also is necessarily accompanied by non-zero fluctuations in the output work.
This further translates to a relation for the thermodynamic efficiency,
\begin{align}
\eta_\text{C} - \eta \geq \eta T_\text{c}  \frac{\av{\dot{W}}}{D_{W,\infty}}  \label{work-bound} ,
\end{align}
which has been previously obtained by Pietzonka and Seifert \cite{Pie18}.
Obviously, the thermodynamic efficiency cannot exceed the Carnot efficiency.
Moreover, as a consequence of the bound \eqref{work-bound}, attaining Carnot efficiency is only possible if the ratio of the average power and the dispersion of work vanishes.
This is similar to the tradeoff relation between power and efficiency derived in Ref.~\cite{Shi16}, where it was concluded that the Carnot efficiency can only be realized at zero power.
The bound \eqref{work-bound} suggests that Carnot efficiency might in principle be reached at finite power, provided that the variance of the work diverges---which of course renders the system useless as an engine.
The same connection between Carnot efficiency at finite power and diverging work fluctuations was recently observed in the context of a heat engine operating close to a critical point in Ref.~\cite{Hol17}.

\subsection{Entropic currents}
\begin{figure}
\includegraphics[width=.47\textwidth]{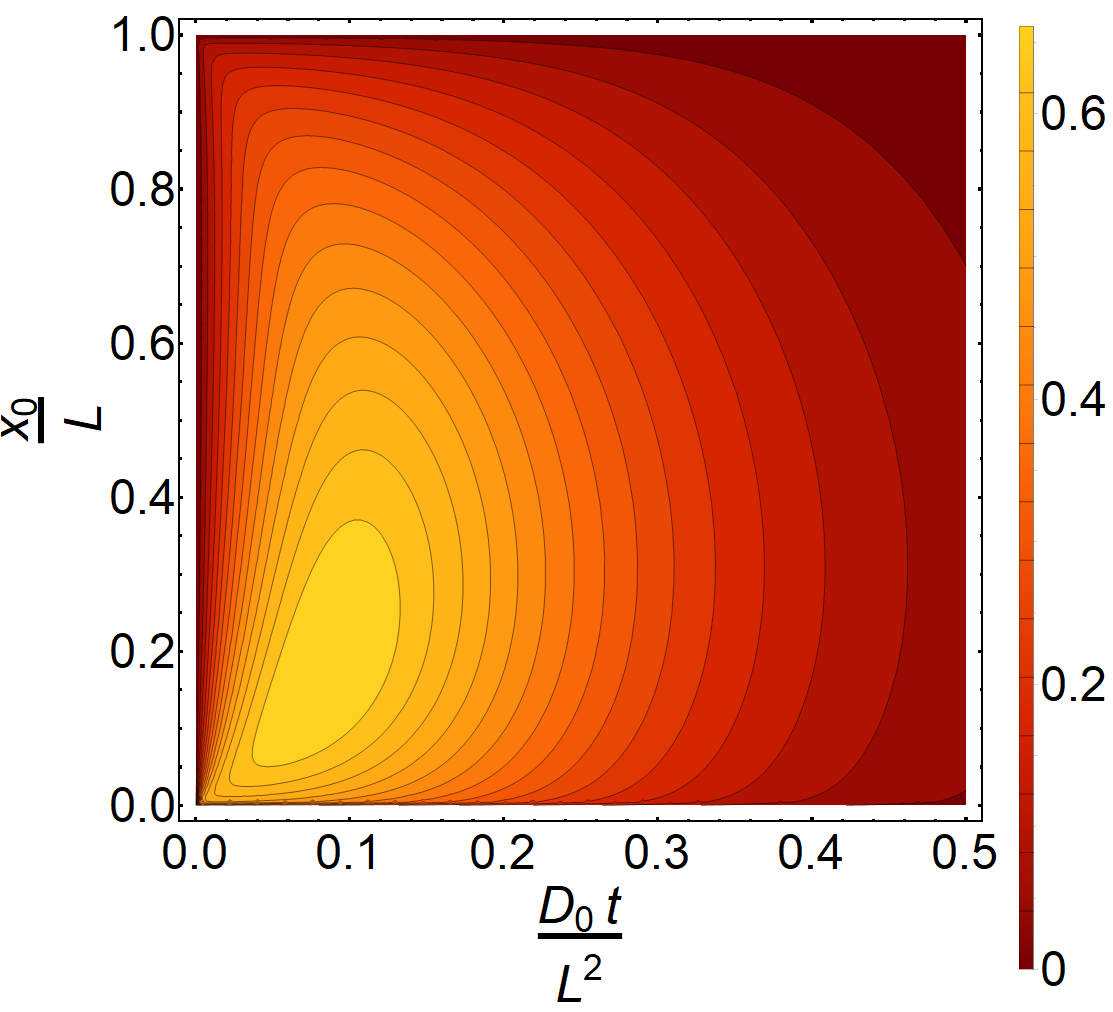}
\caption{(Color online) The ratio between the magnitude of the current $|\langle \dot{x} \rangle|$ and the upper bound obtained by estimating the right hand side of Eq.~\eqref{current-bound-entropic} for a particle diffusing in a one-dimensional box of length $L$. The vertical axis is the width of the initial uniform distribution relative to the size of the box, the horizontal axis is the reduced time, where $D_0$ is the free-space diffusion coefficient. \label{fig-2}}
\end{figure}

So far, we have only considered examples of non-equilibrium steady state systems, where the current is sustained by a heat flow.
However, even in the absence of any heat flow, a transient current may flow if the system is prepared in an out-of-equilibrium initial state.
In this case, the entropy production is given by the Shannon entropy difference between the final and initial state, $\bar{\sigma}_\mathcal{T} = \Delta S(\mathcal{T})/\mathcal{T}$ and thus
\begin{align}
\av{\dot{R}}_\mathcal{T}^2 \leq \frac{\av{\Delta R^2}_\mathcal{T} \Delta S(\mathcal{T})}{2 \mathcal{T}^2} \label{current-bound-entropic} .
\end{align}
The bound on the transport efficiency Eq.~\eqref{transport-efficiency} thus remains valid for such a purely entropy-driven transport.
The bound \eqref{current-bound-entropic} in terms of the Shannon entropy holds whenever there is no net heat exchange between the system and the heat bath.
In the presence of an average heat exchange $\av{Q} \neq 0$ with the bath at temperature $T$, $\Delta S(\mathcal{T})$ has to be replaced by the total entropy production $\Delta S(\mathcal{T}) - \av{Q}_\mathcal{T}/T$.
As a straightforward, nevertheless important, application, we consider a Brownian particle in a bounded domain of volume $V$.
To exclude the possibility of any heat exchange with the bath, we assume the boundaries to consist of hard walls.
In this case, the mean-square fluctuations of the particle displacement are obviously bounded by the maximum distance between any two points of the domain squared, $\av{\Delta x^2}_\mathcal{T} \leq L_\text{max}^2$. 
Similarly, the maximum entropy production is the Shannon entropy difference between the initial state and the equilibrium state, giving a very general bound on any entropic particle current resulting from diffusive spreading of particles in a bounded domain.
Since the Shannon entropy in the uniform final state is essentially the logarithm of the volume, the right hand side of Eq.~\eqref{current-bound-entropic} can be estimated using only on the initial state and the geometry of the domain.
For the simplest example, a particle in a one-dimensional box of length $L$ starting from an initial uniform distribution over a length $x_0$ next to one of the edges of the box, the Shannon entropy difference is $\Delta S_\text{max} = \ln (L/x_0)$.
We can further estimate the mean-square fluctuations of the particle displacement by $\av{\Delta x^2}_\mathcal{T} \leq (L^2 + x_0^2)/12$, which is the sum of the position variance in the final and initial state.
In this case, the current can also be computed analytically (see Appendix \ref{app-rel} for details), allowing for comparison between the current and the estimated bound Eq.~\eqref{current-bound-entropic}.
The result is shown in Fig.~\ref{fig-2}.
At short times, the estimate of the upper bound diverges, while at long times the current decays exponentially, and there is a large discrepancy between the estimate and the actual current.
However, for intermediate times, the current can reach more than 60\% of the estimated bound.

\section{Proof of the bound on the generating function}
The main idea of the proof is to express the scaled cumulant generating function as a path integral,
\begin{align}
K_R(h,\mathcal{T}) = \frac{1}{\mathcal{T}} \ln \bigg( \int \mathcal{D}\Gamma \ e^{h \int_0^\mathcal{T} \text{d}t \ \dot{R}(t)} \mathcal{P}(\Gamma) \bigg) ,
\end{align}
where $\mathcal{P}(\Gamma)$ is the probability of observing a trajectory $\Gamma = \lbrace\bm{x}(t)\rbrace_{0 \leq t \leq \mathcal{T}}$,
which is given by
\begin{align}
\mathcal{P}(\Gamma) = \mathcal{Z}^{-1} &\exp \bigg[ - \int_0^\mathcal{T} \text{d}t \ \mathcal{S}(\bm{x}(t),t) \bigg] \qquad \text{with} \\
\mathcal{S}(\bm{x}(t),t) &= \frac{1}{4} \Big( \dot{\bm{x}}(t) - \bm{A}(t) \Big) \bm{B}^{-1}(t) \Big( \dot{\bm{x}}(t) - \bm{A}(t) \Big)  \n .
\end{align}
We stress that such a continuous-time path integral form should be interpreted as a shorthand for a time-discrete path $\Gamma = \lbrace\bm{x}^N, \bm{x}^{N-1}, \ldots, \bm{x}^0\rbrace$ with $\bm{x}^k = \bm{x}(k \tau) = \bm{x}(k \mathcal{T}/N)$ for sufficiently large $N$.
We employ a time-forward discretization $\dot{\bm{x}}^k = (\bm{x}^k - \bm{x}^{k-1})/\tau$ and $\bm{A}^k = \bm{A}(\bm{x}^{k-1},(k-1)\tau)$, however, any other discretization leads to the same propagator and thus the same result \cite{Wis79}.
In terms of the modified dynamics Eq.~\eqref{fokkerplanck-mod}, we can rewrite the action functional $\mathcal{S}$ as
\begin{align}
\mathcal{S} &= \frac{1}{4} \bigg( \Big( \dot{\bm{x}} - \bm{A} - \bm{Y} \Big) \bm{B}^{-1} \Big( \dot{\bm{x}} - \bm{A} - \bm{Y} \Big) + \bm{Y} \bm{B}^{-1} \bm{Y} + 2 \bm{Y} \bm{B}^{-1} \Big( \dot{\bm{x}} - \bm{A} - \bm{Y} \Big)   \bigg) ,
\end{align}
omitting the dependence on $t$ for brevity.
We identify the first term as the action functional $\mathcal{S}^Y$ in the path probability density of the modified dynamics.
Since the normalization factor $\mathcal{Z}$ does not depend on the drift coefficient and is thus the same for the original and modified dynamics, we can write the cumulant generating function as
\begin{align}
&K_R(h,\mathcal{T}) = \frac{1}{\mathcal{T}} \ln \Bigg( \int \mathcal{D}\Gamma \ \exp\Bigg[\int_0^\mathcal{T} \text{d}t \ \bigg(h \dot{R} \label{cgf-mod}  - \frac{1}{4} \bigg(  \bm{Y} \bm{B}^{-1} \bm{Y} + 2 \bm{Y} \bm{B}^{-1} \Big( \dot{\bm{x}} - \bm{A} - \bm{Y} \Big) \bigg) \bigg) \Bigg] \mathcal{P}^Y(\Gamma) \Bigg) .
\end{align}
We now apply Jensen's inequality to the logarithm to find the lower bound
\begin{align}
&K_R(h,\mathcal{T}) \geq \frac{1}{\mathcal{T}}\int \mathcal{D}\Gamma \ \Bigg(\int_0^\mathcal{T} \text{d}t \ \bigg(h \dot{R} - \frac{1}{4} \bigg(  \bm{Y} \bm{B}^{-1} \bm{Y} + 2 \bm{Y} \bm{B}^{-1} \Big( \dot{\bm{x}} - \bm{A} - \bm{Y} \Big) \bigg) \bigg) \Bigg) \mathcal{P}^Y(\Gamma)  .
\end{align}
The last term can be written as $2 \bm{Y} \bm{B}^{-1} \cdot \bm{\xi}$, which averages to zero due to the non-anticipating It{\=o}-product.
Thus,
\begin{align}
K_R(h,&\mathcal{T}) \geq \frac{1}{\mathcal{T}} \int_0^\mathcal{T} \text{d}t \Big( h \av{\dot{R}}^Y - \frac{1}{4} \av{ \bm{Y} \bm{B}^{-1} \bm{Y}}^Y \Big) \label{variational-2} ,
\end{align}
which is the same as Eq.~\eqref{variational}.
We stress that this result is valid for arbitrary time-dependent drift and diffusion coefficients.
Note that formally, the introduction of the additional generalized forces $\bm{Y}(\bm{x},t)$ corresponds to an absolutely continuous transformation of the stochastic process \cite{Gir60}.
Expressing the generating function in terms of the path probability of the transformed process can then be accomplished by introducing the Radon-Nikodym derivative, see e.~g.~Refs.~\cite{Har12,Che15b}.
Applying Jensen's inequality then yields the second term on the right-hand side of Eq.~\eqref{variational-2} as the relative entropy between the two path measures \cite{Gir60,Har12,Che15b}.

\section{Discussion}
We have derived a bound on the cumulant generating function of a generalized current, valid for general Langevin dynamics driven by Gaussian white noise.
In contrast to previous studies \cite{Gin16,Hor17}, our proof of the bound does not rely on large deviation arguments, but follows directly from the path integral representation of the generating function.
Consequently, it holds out of steady state and for time-dependent driving and currents.

Without explicit time-dependence, we obtain the quadratic bound Eq.~\eqref{cgf-bound} for the generating function, which implies the thermodynamic uncertainty relation Eq.~\eqref{uncertainty}.
The transport efficiency Eq.~\eqref{transport-efficiency} derived from the uncertainty relation measures how close a given system comes to saturating the bound.
Its applicability to arbitrary currents in both steady state and relaxational dynamics allows for a wide range of systems to be classified according to their ability to provide reliable transport while minimizing dissipation. 
We thus expect this efficiency to be a useful benchmark for stochastic transport, both in theoretical models and applications, for example to active Brownian motion or molecular motors.

Finally, the bound on entropic currents in terms of the Shannon entropy difference lends further support to the physical relevance of the latter in non-equilibrium situations \cite{Leb93,Gol04,Gav17}.
While the Shannon entropy is usually not easily accessible in experiments, one may use Eq.~(17) to obtain a lower bound on the former,
\begin{align}
\Delta S(\mathcal{T}) \geq \frac{2 \av{R}_\mathcal{T}^2}{\av{\Delta R^2}_\mathcal{T}} + \frac{\av{\Delta Q}_\mathcal{T}}{T}.
\end{align}
Thus, the Shannon entropy change is bounded from below by a combination of measurable observables.

\acknowledgments
{The present study was supported by KAKENHI (Nos. 25103002, 17H01148 and 15F15324).
A.~D.~was employed as an International Research Fellow of the Japan Society for the Promotion of Science.
The authors wish to thank N. Shiraishi, R. Chetrite and M. Rosinberg for stimulating discussions.}

\appendix

\section{Optimal generalized forces} \label{app-optimal}

\setcounter{equation}{0}

In the main text, we derived the bound on the cumulant generating function of a generalized current
\begin{align}
K_R(h,\mathcal{T}) \geq \frac{1}{\mathcal{T}}\int_0^\mathcal{T} \text{d}t \ \bigg( h \av{\dot{R}}_t^Y - \frac{1}{4} \av{\bm{Y} \bm{B}^{-1}\bm{Y}}_t^Y \bigg) \label{bound} ,
\end{align} 
with the averages being taken with respect to the probability density $P^Y$ determined by
\begin{align}
\partial_t P^Y(\bm{x},&t) = -\bm{\nabla} \bm{J}^Y(\bm{x},t)  \quad \text{with} \quad
\bm{J}^Y(\bm{x},t) = \Big(\bm{A}(\bm{x},t) + \bm{Y}(\bm{x},t)  - \bm{\nabla} \bm{B}(\bm{x},t) \Big) P^Y(\bm{x},t) \label{fp-mod},
\end{align}
with initial condition $P^Y(\bm{x},0) = P_0(\bm{x})$.
We now want to find the set of generalized forces $\bm{Y}(\bm{x},t)$ that maximize the bound.
To do so, we first allow the modified dynamics to start from an arbitrary initial condition $P^Y(\bm{x},0) = P^Y_0(\bm{x})$.
This leads to an additional term in Eq.~\eqref{bound},
\begin{align}
K_R(h,\mathcal{T}) \geq \frac{1}{\mathcal{T}}\int_0^\mathcal{T} \text{d}t \ \bigg( h \av{\dot{R}}_t^Y - \frac{1}{4} \av{\bm{Y} \bm{B}^{-1}\bm{Y}}_t^Y \bigg) - \frac{1}{\mathcal{T}}\int \text{d}\bm{x} \ P^Y_0(\bm{x}) \log \bigg(\frac{P^Y_0(\bm{x})}{P_0(\bm{x})} \bigg) \label{bound-initial},
\end{align}
which corresponds to minus the relative entropy between the two initial conditions and is thus always negative.
Next, we maximize the functional
\begin{align}
\tilde{\Psi}^Y(h,\mathcal{T}) &= \frac{1}{\mathcal{T}} \int_0^\mathcal{T} \text{d}t \int \text{d}\bm{x} \ \psi \Big(P^Y,\partial_t P^Y ,\bm{\nabla} P^Y ,\bm{J}^Y, \bm{\nabla}\bm{J}^Y, \bm{Y},\alpha,\bm{\beta},\bm{x},t \Big) \\
\text{with} \quad \psi &= h \bm{Z}\bm{J}^Y - \frac{1}{4} \bm{Y}\bm{B}^{-1}\bm{Y} P^Y - \alpha \Big(\partial_t P^Y + \bm{\nabla}\bm{J}^Y \Big) - \bm{\beta}  \Big(\bm{J}^Y - \big( \bm{A} + \bm{Y} - \bm{\nabla}\bm{B} \big) P^Y \Big) \n ,
\end{align}
where we included the constraints Eq.~\eqref{fp-mod} with the Lagrange multipliers $\alpha(\bm{x},t)$ and $\bm{\beta}(\bm{x},t)$.
The Euler Lagrange equations read
\begin{subequations}
\begin{align}
\frac{\partial \psi}{\partial P^Y} - \frac{\text{d}}{\text{d}t} \frac{\partial \psi}{\partial (\partial_t P^Y)} - \sum_i \partial_{x_i} \frac{\partial \psi}{\partial (\partial_{x_i} P^Y)}  &= - \frac{1}{4}\bm{Y}\bm{B}^{-1}\bm{Y} + \bm{\beta} \big(\bm{A}+\bm{Y} - \bm{\nabla}\bm{B} \big) + \partial_t \alpha + \bm{\nabla}\big(\bm{\beta}\bm{B}\big) = 0 \\
\frac{\partial \psi}{\partial J_i^Y} - \sum_i \partial_{x_i} \frac{\partial \psi}{\partial (\partial_{x_i} J_i^Y)} &= h Z_i - \beta_i+ \partial_{x_i} \alpha = 0 \\
\frac{\partial \psi}{\partial Y_i} &= \frac{1}{2} \big(\bm{B}^{-1} \bm{Y} \big)_i P^Y + \beta_i P^Y = 0 ,
\end{align}
\end{subequations}
together with Eq.~\eqref{fp-mod}.
In addition, since $P^Y(\bm{x},T)$ is unconstrained, we have to add the transversality condition
\begin{align}
\frac{\partial \psi}{\partial (\partial_t P^Y)} \bigg\vert_{t = \mathcal{T}} = -\alpha_\mathcal{T}(\bm{x},\mathcal{T}) = 0 \label{transversal}.
\end{align}
Here we added the subscript $\mathcal{T}$ to the function $\alpha_\mathcal{T}$ to indicate that the latter depends on the total time $\mathcal{T}$ via the above terminal condition.
These equations can be solved for the optimal force $\bm{Y}^*$,
\begin{align}
\bm{Y}^*(\bm{x},t) = 2 \bm{B}(\bm{x},t) \Big( h \bm{Z}(\bm{x},t) + \bm{\nabla} \alpha_\mathcal{T}(\bm{x},t) \Big) \label{force-optimal} ,
\end{align}
where $\alpha_\mathcal{T}$ has to solve the partial differential equation
\begin{align}
-\partial_t \alpha_\mathcal{T}(\bm{x},t) &= \Big[ h \bm{Z}(\bm{x},t) + \bm{\nabla} \alpha_\mathcal{T}(\bm{x},t) \Big]\bm{B}(\bm{x},t) \Big[ h \bm{Z}(\bm{x},t) + \bm{\nabla} \alpha_\mathcal{T}(\bm{x},t) \Big] \label{alpha-eq}  \\
& \hspace{3cm} + \Big(\bm{A}(\bm{x},t) + \bm{B}(\bm{x},t)\bm{\nabla}\Big)\Big( h \bm{Z}(\bm{x},t) + \bm{\nabla} \alpha_\mathcal{T}(\bm{x},t) \Big) \n 
\end{align}
with the appropriate terminal condition.
Here we use the convention that differential operators act through parentheses, $(\partial_x f) g = g \partial_x f + f \partial_x g$, whereas they act only inside brackets, $[\partial_x f] g = g \partial_x f$.
Using this solution, we can evaluate the optimal value of the bound \eqref{bound-initial},
\begin{align}
\Psi^*(h,\mathcal{T}) = \frac{1}{\mathcal{T}}\int_0^\mathcal{T} \text{d}t \bigg( h \av{\dot{R}}_t^{Y^*} - \frac{1}{4} \av{\bm{Y}^* \bm{B}^{-1}\bm{Y}^*}_t^{Y^*} \bigg) ,
\end{align}
where we omit the initial term for now.
Using Eq.~\eqref{alpha-eq}, we can replace the second term
\begin{align}
\Psi^*(h,\mathcal{T}) &= \frac{1}{\mathcal{T}}\int_0^\mathcal{T} \text{d}t \int\text{d}\bm{x} \bigg( h \bm{Z}(\bm{x},t).\bm{J}^{Y^*}(\bm{x},t) + \partial_t \alpha_\mathcal{T}(\bm{x},t) P^{Y^*}(\bm{x},t) \\
& \qquad \qquad \qquad + \Big[ \Big(\bm{A}(\bm{x},t) + \bm{B}(\bm{x},t)\bm{\nabla}\Big)\Big( h \bm{Z}(\bm{x},t) + \bm{\nabla} \alpha_\mathcal{T}(\bm{x},t) \Big) \Big] P^{Y^*}(\bm{x},t) \bigg) \n .
\end{align}
We now integrate by parts and add and subtract a term
\begin{align}
\Psi^*(h,\mathcal{T}) &= \frac{1}{\mathcal{T}}\int_0^\mathcal{T} \text{d}t \int\text{d}\bm{x} \bigg( h \bm{Z}(\bm{x},t).\bm{J}^{Y^*}(\bm{x},t) + \partial_t \alpha_\mathcal{T}(\bm{x},t) P^{Y^*}(\bm{x},t) \nn
& \qquad \qquad  + \Big[ h \bm{Z}(\bm{x},t) + \bm{\nabla} \alpha_\mathcal{T}(\bm{x},t) \Big]\Big(\bm{A}(\bm{x},t) + \bm{Y}^*(\bm{x},t) - \bm{\nabla}\bm{B}(\bm{x},t)\Big) P^{Y^*}(\bm{x},t) \nn
& \qquad \qquad - \Big[ h \bm{Z}(\bm{x},t) + \bm{\nabla} \alpha_\mathcal{T}(\bm{x},t) \Big]\bm{Y}^*(\bm{x},t) P^{Y^*}(\bm{x},t) \bigg) \nn
&= \frac{1}{\mathcal{T}}\int_0^\mathcal{T} \text{d}t \int\text{d}\bm{x} \bigg( 2 h \bm{Z}(\bm{x},t)\bm{J}^{Y^*}(\bm{x},t) - \frac{1}{2} \bm{Y}^*(\bm{x},t) \bm{B}^{-1}(\bm{x},t) \bm{Y}^*(\bm{x},t) P^{Y^*}(\bm{x},t) \nn
& \qquad \qquad + \partial_t \alpha_\mathcal{T}(\bm{x},t) P^{Y^*}(\bm{x},t)  + \Big[\bm{\nabla} \alpha_\mathcal{T}(\bm{x},t) \Big]\bm{J}^{Y^*}(\bm{x},t) \bigg)  .
\end{align}
The first term on the right hand side is precisely $2 \Psi^*$.
Integrating by parts in the last term, we get
\begin{align}
\Psi^*(h,\mathcal{T}) &= 2 \Psi^*(h,\mathcal{T}) + \frac{1}{\mathcal{T}}\int_0^\mathcal{T} \text{d}t \int\text{d}\bm{x}   \bigg( \partial_t \alpha_\mathcal{T}(\bm{x},t) P^{Y^*}(\bm{x},t)  -  \alpha_\mathcal{T}(\bm{x},t) \bm{\nabla}\bm{J}^{Y^*}(\bm{x},t) \bigg)  .
\end{align}
Since the gradient of the probability current is minus the derivative of the probability density via Eq.~\eqref{fp-mod}, the second term is a total time-derivative and we get
\begin{align}
\Psi^*(h,\mathcal{T}) &= - \frac{1}{\mathcal{T}} \int \text{d}\bm{x} \ \Big( \alpha_{\mathcal{T}}(\bm{x},\mathcal{T}) P^{Y^*}(\bm{x},\mathcal{T}) - \alpha_{\mathcal{T}}(\bm{x},0) P^{Y}_0(\bm{x}) \Big) .
\end{align}
Using the terminal condition $\alpha_\mathcal{T}(\bm{x},\mathcal{T}) = 0$, and taking into account the relative entropy between the initial conditions, we arrive at
\begin{align}
K_R(h,\mathcal{T}) \geq \frac{1}{\mathcal{T}} \int\text{d}\bm{x} \ \alpha_{\mathcal{T}}(\bm{x},0) P^{Y}_0(\bm{x}) - \frac{1}{\mathcal{T}}\int \text{d}\bm{x} \ P^Y_0(\bm{x}) \log \bigg(\frac{P^Y_0(\bm{x})}{P_0(\bm{x})} \bigg)  .
\end{align}
Finally, we maximize this expression with respect to $P^Y_0(\bm{x})$, leading to
\begin{align}
P^Y_0(\bm{x}) = \frac{e^{\alpha_\mathcal{T}(\bm{x},t)} P_0(\bm{x})}{\int \text{d}\bm{x} \ e^{\alpha_\mathcal{T}(\bm{x},t)} P_0(\bm{x})} ,
\end{align}
and thus the bound
\begin{align}
K_R(h,\mathcal{T}) \geq \frac{1}{\mathcal{T}} \ln \bigg( \int \text{d}\bm{x} \ e^{\alpha_\mathcal{T}(\bm{x},0)} P_0(\bm{x}) \bigg). \label{bound-optimal}
\end{align}
On the other hand, we can use the expression for the generating function (see Eq.~(21) of the main text),
\begin{align}
K_R(h,\mathcal{T}) = \frac{1}{\mathcal{T}} \ln \Bigg( &\int \mathcal{D}\Gamma \ \exp\Bigg[\int_0^\mathcal{T} \text{d}t \ \bigg(h \dot{R} \\
& - \frac{1}{4} \bigg(  \bm{Y} \bm{B}^{-1} \bm{Y} - 2 \bm{Y} \bm{B}^{-1} \Big( \dot{\bm{x}} - \bm{A} - \bm{Y} \Big) \bigg) \bigg) \Bigg] \mathcal{P}^Y(\Gamma) \frac{P_0(\bm{x^0})}{P^Y_0(\bm{x^0})} \Bigg) \n ,
\end{align}
and evaluate the exponent for the optimal force,
\begin{align}
h &\dot{R}(t) - \frac{1}{4} \bigg(  \bm{Y}^*(t) \bm{B}^{-1}(t) \bm{Y}^*(t) - 2 \bm{Y}^*(t) \bm{B}^{-1}(t) \Big( \dot{\bm{x}}(t) - \bm{A}(t) - \bm{Y}^*(t) \Big) \bigg) \nn
&= h \Big( \bm{Z}(t) \cdot \dot{\bm{x}}(t) + \big(\bm{B}(t)\bm{\nabla}\big)\bm{Z}(t) \Big) + \Big[h \bm{Z}(t) + \bm{\nabla}\alpha(t) \Big]\bm{B}(t) \Big[h \bm{Z}(t) + \bm{\nabla}\alpha(t) \Big] - \Big[h \bm{Z}(t) + \bm{\nabla}\alpha(t) \Big] \Big(\dot{\bm{x}}(t)-\bm{A}(t)\Big) \nn
&= h \big(\bm{B}(t)\bm{\nabla}\big)\bm{Z}(t) - \partial_t \alpha(t) - \Big(\bm{A}(t) + \bm{B}(t)\bm{\nabla} \Big)\Big(h\bm{Z}(t) + \bm{\nabla}\alpha(t) \Big) - h \bm{Z}(t)\bm{A}(t) - \bm{\nabla}\alpha(t) \cdot \dot{\bm{x}}(t) + \bm{A}(t) \bm{\nabla}\alpha(t)\Big) \nn
&= - \partial_t \alpha(t) - \bm{\nabla}\alpha(t) \cdot \dot{\bm{x}}(t) - \big(\bm{B}(t)\bm{\nabla}\big)\bm{\nabla}\alpha(t),
\end{align}
where we used the evolution equation for $\alpha$, Eq.~\eqref{alpha-eq}, from the second to the third line.
Applying the It{\=o}-formula for $\alpha_\mathcal{T}(\bm{x}(t),t)$,
\begin{align}
\frac{\text{d}}{\text{d}t} \alpha_\mathcal{T}(\bm{x}(t),t) = \partial_t \alpha_\mathcal{T}(\bm{x}(t),t) + \bm{\nabla} \alpha_\mathcal{T}(\bm{x}(t),t) \cdot \dot{\bm{x}}(t) + \big(\bm{B}(\bm{x}(t),t)\bm{\nabla}\big)\bm{\nabla}\alpha_\mathcal{T}(\bm{x}(t),t),
\end{align}
the exponent becomes a total derivative and we can write
\begin{align}
&K_R(h,\mathcal{T}) = \frac{1}{\mathcal{T}} \ln \Bigg( \int \mathcal{D}\Gamma \ \exp\Big[\alpha_\mathcal{T}(0)-\alpha_\mathcal{T}(\mathcal{T})\Big] \mathcal{P}^{Y^*}(\Gamma) \frac{P_0(\bm{x^0})}{P^Y_0(\bm{x^0})} \Bigg) .
\end{align}
Since $\alpha_\mathcal{T}(\bm{x},\mathcal{T}) = 0$ via Eq.~\eqref{transversal}, the path integral contracts to the integral over the initial density
\begin{align}
K_R(h,\mathcal{T}) = \frac{1}{\mathcal{T}} \ln \bigg( \int \text{d}\bm{x} \ e^{\alpha_\mathcal{T}(\bm{x},0)} P_0(\bm{x}) \bigg).
\end{align}
This is exactly the same as the bound Eq.~\eqref{bound-optimal}.
Thus the above choice for the optimal generalized forces, Eq.~\eqref{force-optimal}, not only maximizes the bound but actually leads to equality between the generating function and the bound.
This proves that, provided one can find an appropriate solution to Eq.~\eqref{alpha-eq}, there exists a set of generalized forces which leads to equality in Eq.~\eqref{bound-initial}, permitting to write the variational formula
\begin{align}
K_R(h,\mathcal{T}) = \frac{1}{\mathcal{T}} \underset{{\bm{Y},P_0^Y}}{\text{sup}} \Bigg[ \int_0^\mathcal{T} \text{d}t \ \bigg( h \av{\dot{R}}_t^Y - \frac{1}{4} \av{\bm{Y} \bm{B}^{-1}\bm{Y}}_t^Y \bigg) - \int \text{d}\bm{x} \ P^Y_0(\bm{x}) \log \bigg(\frac{P^Y_0(\bm{x})}{P_0(\bm{x})} \bigg) \Bigg],
\end{align}
where the maximization is taken over all added generalized forces $\bm{Y}(\bm{x},t)$ and all initial conditions $P^Y_0(\bm{x})$.
This variational formula is the generalization of the result derived by Nemoto et al.~\cite{Nem11} to arbitrary time-dependent dynamics at finite time.
At $h=0$, we obviously have $\alpha_\mathcal{T}(\bm{x},t) = 0$.
Assuming that we can expand $\alpha_\mathcal{T}$ as a power series in $h$, we express the cumulants $\kappa_R^{(n)}(\mathcal{T})$ of the current via derivatives of $\alpha_\mathcal{T}$,
\begin{align}
\kappa_R^{(n)}(\mathcal{T}) = \int \text{d}\bm{x} \ \partial_h^{(n)} \alpha_\mathcal{T}(\bm{x},0) \Big\vert_{h = 0} P_0(\bm{x}) .
\end{align}

\section{Current, diffusivity and entropy for the B{\"u}ttiker-Landauer ratchet} \label{app-blr}

\setcounter{equation}{0}

The Fokker-Planck equation corresponding to the Langevin equation is
\begin{align}
\partial_t P(x,t) = \frac{1}{\gamma} \partial_x \Big( U'(x) + T(x) \partial_x \Big) P(x,t) .
\end{align}
We want to find the periodic stationary solution $P^\text{st}(x+L) = P^\text{st}(x)$ to this equation.
The stationary probability current is constant in the one-dimensional case,
\begin{align}
J^\text{st} = -\frac{1}{\gamma} \Big[ U'(x) + T(x) \partial_x \Big] P^\text{st}(x) .
\end{align}
We may solve this for $P^\text{st}(x)$ to obtain the general solution
\begin{align}
P^\text{st}(x) = \frac{e^{-\psi(x)}}{T(x)} \bigg[ C_1 - \gamma J^\text{st} \int_0^x \text{d}y \ e^{\psi(y)} \bigg] \qquad \text{with} \qquad \psi(x) = \int_0^x \text{d}y \ \frac{U(y)}{T(y)} .
\end{align}
The constants $C_1$ and $J^\text{st}$ are determined from periodicity and the normalization condition $\int_0^L \text{d}x \ P^\text{st}(x) = 1$ \cite{Bue87},
\begin{align}
P^\text{st}(x) = \gamma \frac{J^\text{st}}{1-e^{\psi(L)}} \frac{e^{-\psi(x)}}{T(x)} \int_x^{x+L} \text{d}y \ e^{\psi(y)} \quad \text{with} \quad
J^\text{st} = \frac{1}{\gamma} \frac{1-e^{\psi(L)}}{\int_0^L \text{d}x \ e^{\psi(x)} \int_{x-L}^x \text{d}y \ \frac{e^{-\psi(y)}}{T(y)}} .
\end{align}
The probability current is related to the steady state velocity via $v^\text{st} = L J^\text{st}$.
For the long-time diffusivity of a particle in a periodic potential, Reimann et al. \cite{Rei01} derived an exact result in terms of the moments of the first passage time $\Theta(0 \rightarrow L)$ of a particle starting at $x=0$ to reach $x=L$,
\begin{align}
D_{\infty} = \frac{L^2}{2} \frac{\av{\Theta^2(0 \rightarrow L)}- \av{\Theta(0 \rightarrow L)}^2}{\av{\Theta(0 \rightarrow L)}^3}.
\end{align} 
The corresponding expressions for a position-dependent mobility $\gamma(x)$ have been obtained by Krishnan et al. \cite{Kri05}; the case of a position-dependent temperature follows by analogy,
\begin{align}
D_{\infty} = \frac{L^2}{\gamma} \frac{\int_0^L \text{d}x \ e^{\psi(x)} \big[ \int_x^{x+L}\text{d}y \ e^{\psi(y)} \big] \big[\int_{x-L}^x \text{d}y \ e^{-\psi(y)}/T(y) \big]^2}{\big[\int_0^L \text{d}x \ e^{\psi(x)} \int_{x-L}^x \text{d}y \ e^{-\psi(y)}/T(y)\big]^3} .
\end{align}
Finally, the rate of total entropy production may be written as \cite{Spi12}
\begin{align}
\sigma_\text{t}(t) = \int_0^L \text{d}x \ \frac{J(x,t)^2}{T(x) P(x,t)}.
\end{align}
Since the Shannon entropy production rate $\sigma_\text{s} = -\partial_t \int \text{d}x \ \ln(P(x,t)) P(x,t)$ is zero in the steady state, the entire entropy production is the entropy production in the medium $\sigma_\text{t} = \sigma_\text{m}$, which yields
\begin{align}
\sigma_\text{m} =\frac{\big(1-e^{\psi(L)}\big)^2}{\gamma} \frac{\int_0^L \text{d}x \ e^{\psi(x)} \big[\int_x^{x+L} \text{d}y \ e^{\psi(y)}\big]^{-1}}{\int_0^L \text{d}x \ e^{\psi(x)} \int_{x-L}^x \text{d}y \ e^{-\psi(y)}/T(y)} .
\end{align}
Note that we may equivalently write the less explicit form
\begin{align}
\sigma_\text{m} &= \int_0^L \text{d}x \ \frac{1}{\gamma T(x)} \Big[ V'(x)^2 - T(x) V''(x) \Big] P^\text{st}(x) \qquad \text{with} \qquad V(x) = U(x) + T(x) \label{sigma-bl}.
\end{align}
This form highlights the analogy to entropy production in the presence of a uniform temperature where $V'(x) = U'(x)$ \cite{Sek10}.
Note that while the entropy production is positive for any finite current, the total heat transfer between the particle and the bath is zero for this system.

\section{Entropic current for diffusive relaxation} \label{app-rel}

\setcounter{equation}{0}

As introduced in the main text, we consider a simple example for a diffusive relaxation process.
For a particle in a one-dimensional box of size $L$, the propagator is obtained by solving the diffusion equation
\begin{align}
\partial_t P(x,t) = D_0 \partial_{x}^2 P(x,t),
\end{align}
with the Neumann-type boundary conditions
\begin{align}
\partial_x P(x,t) \Big\vert_{x=0} = \partial_x P(x,t) \Big\vert_{x=L} = 0.
\end{align}
The general solution to the diffusion equation can be written as
\begin{align}
P(x,t) = \Big(a_k \sin(k x) + b_k \cos(k x)\Big) e^{-D_0 k^2 t},
\end{align}
where $a_k$, $b_k$ and $k$ are determined by the initial and boundary conditions.
For an initial density profile $P_0(x)$, the solution reads
\begin{align}
P(x,t) = \frac{1}{L} + \frac{2}{L} \sum_{n = 1}^{\infty} b_n \cos\bigg(\frac{\pi n x}{L} \bigg) e^{-\left(\frac{\pi n}{L}\right)^2 D_0 t} \qquad \text{with} \qquad b_n = \int_0^L \text{d}x \ P_0(x) \cos\bigg(\frac{\pi n x}{L} \bigg) .
\end{align}
From this, we find for the current
\begin{align}
\av{\dot{x}}_t = \int_0^L \text{d}x \ x \partial_t P(x,t) = \frac{4 D_0}{L} \sum_{n=0}^\infty b_{2 n + 1} e^{-\left(\frac{\pi (2n+1)}{L}\right)^2 D_0 t}.
\end{align}
For the particular case when the initial density is uniform
\begin{align}
P_0(x) = \left\lbrace \begin{array}{ll}
\frac{1}{x_0} &\text{for} \quad 0 \leq x \leq x_0 \\
0 &\text{for} \quad x_0 < x \leq 1 ,
\end{array} \right.
\end{align}
the coefficients $b_n$ are given by
\begin{align}
b_n = \frac{L}{\pi n x_0} \sin\bigg(\frac{\pi n x_0}{L} \bigg) .
\end{align}
Then the current can be written in terms of the variables $\xi = x_0/L$ and $\theta = D_0 t/L^2$
\begin{align}
\av{\dot{x}}_t = \frac{4 D_0}{\pi L \xi} \sum_{n=0}^\infty \frac{\sin(\pi (2 n+1) \xi)}{2n+1} e^{-\pi^2 (2n+1)^2 \theta}.
\end{align}
By estimating $\av{\Delta x^2}_t \leq (L^2 + x_0^2)/12$ and
\begin{align}
\Delta S(t) \leq S_\text{infty} - S_0 = \ln \frac{L}{x_0},
\end{align}
we have the bound
\begin{align}
\av{\dot{x}}_t^2 \leq \frac{\av{\Delta x^2}_t \Delta S(t)}{2 t^2} \leq \frac{(L^2 + x_0^2) \ln \frac{L}{x_0}}{24 t^2} \equiv \av{\dot{x}}^2_\text{bound}
\end{align}
We can write the ratio between the current and the estimated bound as
\begin{align}
\bigg\vert \frac{\av{\dot{x}}_t}{\av{\dot{x}}_\text{bound}} \bigg\vert = \frac{16 \sqrt{2} \ \theta}{\pi \xi \sqrt{- (1 + \xi^2)\ln \xi }} \bigg\vert \sum_{n=0}^\infty \frac{\sin(\pi (2 n+1) \xi)}{2n+1} e^{-\pi^2 (2n+1)^2 \theta} \bigg\vert,
\end{align}
which is the expression plotted for $n = 50$---yielding good convergence except for very short times---in Fig.~2 of the main text.

\end{document}